\documentclass[a4paper]{article}

\usepackage{INTERSPEECH2022}
\usepackage[dvipsnames]{xcolor}
\usepackage{bm}
\usepackage{amsmath, amsfonts}
\usepackage{caption}
\usepackage{subcaption}
\usepackage{pgfplots, pgfplotstable}
\usepackage{booktabs}
\usepackage{multirow}
\usepackage{hyperref}
\usepackage{comment}
\usepgfplotslibrary{groupplots}
\usetikzlibrary{
    patterns,matrix,positioning
}
\definecolor{RYB1}{RGB}{218,232,252}
\definecolor{RYB2}{RGB}{255,225,225}

\title{SpeechPrompt: An Exploration of Prompt Tuning on Generative Spoken Language Model for Speech Processing Tasks}

\name{Kai-Wei Chang$^1$, Wei-Cheng Tseng$^1$, Shang-Wen Li$^2$, Hung-yi Lee$^1$
\thanks{The authors acknowledge the support of 2022 Eighth Frederick Jelinek Memorial Summer Workshop.}\thanks{$^2$ Work done while working at Amazon Inc., currently at Meta AI.}}

\address{
  $^1$Graduate Institute of Communication Engineering, National Taiwan University, Taiwan\\$^2$Amazon AI, USA}
  
\email{kaiwei.chang.tw@gmail.com, r09942094@ntu.edu.tw, swdanielli@gmail.com, hungyilee@ntu.edu.tw}

\begin{document}
\pgfplotstableread[col sep=&, header=true]{
description & KS & IC
1 & 94.32 & 93.75
2 & 95.03 & 97.57
3 & 95.16 & 97.84
6  & 94.87 & 98.40
30  & 95.07 & 98.66
60  & 94.87 & 98.63
}\exptwo

\pgfplotstableread[col sep=&, header=true]{
param & KS & IC
2000 & 88.34 & {}
4000 & 90.26 & {}
6000 & 90.62 & {}
9000 & 91.85 & {}
22000 & 92.86 & 79.44
52000 & 93.22 & 87.11
103000  & 93.02 &  92.28
155000  & 93.09 &  94.15
257000  & 92.96 &  95.78
}\expfour

\pgfplotstableread[col sep=&, header=true]{
param & KS & IC
26000 & 94.32 & 93.75
52000 & 95.03 & 97.57
78000 & 95.16 & 97.84
150000 & 94.87 & 98.40
750000 & 95.07 & 98.66
1500000 & 94.87 & 98.63
}\expfourDeep

\maketitle

\begin{abstract}
Speech representations learned from Self-supervised learning (SSL) models can benefit various speech processing tasks.
However, utilizing SSL representations usually requires fine-tuning the pre-trained models or designing task-specific downstream models and loss functions, causing much memory usage and human labor.
Recently, prompting in Natural Language Processing (NLP) has been found to be an efficient technique to leverage pre-trained language models (LMs).
Specifically, prompt tuning optimizes a limited number of task-specific parameters with a fixed pre-trained model; as a result, only a small set of parameters is needed to be stored for each task.
Prompt tuning improves computation and memory efficiency by leveraging the pre-trained LM’s prediction ability.
Nevertheless, such a paradigm is little studied in the speech community.
We report in this paper the first exploration of the prompt tuning paradigm for speech processing tasks based on Generative Spoken Language Model (GSLM).
Experiment results show that the prompt tuning technique achieves competitive performance in speech classification tasks with fewer trainable parameters than fine-tuning specialized downstream models.
We further study the technique in challenging sequence generation tasks. Prompt tuning also demonstrates its potential, while the limitation and possible research directions are discussed in this paper.
The source code is available on \url{https://github.com/ga642381/SpeechPrompt}.
\end{abstract}


\section{Introduction}
Recently, SSL speech models \cite{DBLP:journals/corr/abs-1807-03748, DBLP:conf/nips/BaevskiZMA20, DBLP:journals/taslp/HsuBTLSM21, chen2021wavlm} can achieve state-of-the-art performance in a wide variety of speech processing tasks.
The learned, highly informative representations can benefit various speech processing downstream tasks \cite{DBLP:conf/icassp/BaevskiM20, DBLP:journals/corr/abs-2111-09296, DBLP:conf/icassp/LaiCL0G21, DBLP:conf/interspeech/LinLCL21}.
Despite their success, adopting these models to different downstream tasks requires either (a) fine-tuning the pre-trained model \cite{DBLP:conf/nips/BaevskiZMA20, DBLP:journals/taslp/HsuBTLSM21, baevski2021unsupervised} or (b) appropriately designing downstream models and loss functions \cite{DBLP:conf/interspeech/YangCCLLLLSCLHT21, Tsai2022SUPERBSGES}, which results in an increasing burden of human labor and memory cost \cite{lai2021parp} as the number of tasks scales \cite{Tsai2022SUPERBSGES}. Hence, there is a demand for exploring an alternative paradigm to leverage SSL Speech models.

On the other hand, in the Natural Language Processing (NLP) field, prompting methods have gained researchers' attention \cite{DBLP:journals/corr/abs-2107-13586}. The methods scale up pre-trained language models (LMs) at serving multiple downstream tasks in a unified and efficient way.
For each downstream task, prompting methods aim to find task-specific templates or a limited number of parameters that steer LMs to generate results for the task without modifying LM's parameters.
For example, in a sentiment classification task for movie review, we can design a prompt ``[X] The movie is \_\_".
The LM takes a sentence to be classified and fits it into the template at [X].
By generating a sentiment word from a pre-defined set of tokens (e.g. great, neutral, bad) that one-to-one mapped to classification labels, we transform the sentiment classification task into a generation problem.
Alternatively, prompts are not necessary to be readable by humans. 
Researchers proposed prompt tuning methods that learn continuous prompts \cite{DBLP:journals/corr/abs-2107-13586, DBLP:conf/acl/LiL20, DBLP:journals/corr/abs-2103-10385, DBLP:journals/corr/abs-2110-07602} in models' embedding space.
Studies have shown that prompt methods can reformulate most NLP tasks as generation problems and yield competitive performance \cite{DBLP:journals/corr/abs-2107-13586}. 

The prompting paradigm is appealing as the number of downstream tasks to be served increases. 
Rather than requiring a specialized downstream model for each task, a single generalist model can simultaneously serve many different tasks in one inference batch.
Since parameters of tuned prompts are usually several orders smaller than parameters of LMs \cite{DBLP:conf/emnlp/LesterAC21}, the prompting paradigm significantly improves memory and computation efficiency.
Furthermore, there is a unified inference process with the original pre-trained LM for all downstream tasks in the paradigm. 
Hence, less human labor is required in model authoring for each task. 
Despite the success in NLP, there is little research on the prompting paradigm in the speech community.

To bring the benefit of the prompting paradigm to the speech processing field, we propose a prompt tuning framework for multiple downstream speech processing tasks, including Keyword Spotting (KS), Intent Classification (IC), Automatic Speech Recognition (ASR), and Slot Filling (SF). 
The framework unifies training and inference for multiple tasks by leveraging the generation capability of the pre-trained LM.
To our best knowledge, our work is the first study in the prompting paradigm that achieves competitive performance in various speech processing tasks.

We utilize \textbf{Generative Spoken Language Model (GSLM)} \cite{DBLP:journals/corr/abs-2102-01192} as our backbone LM and apply prompting on top of it. GSLM is used, for GSLM is the first generative speech LM pre-trained on a large-scale speech dataset, and it has a large model capacity to generate meaningful output.
Experiment shows that the proposed framework achieves competitive accuracy (Acc) in single-label and multi-label speech classification tasks.
While the framework demonstrates the potential of prompting GSLM, we also identify the limitations when performing challenging sequence generation tasks and discuss the potential research directions in the paper.
We hope by exploring and analyzing the novel prompting paradigm for speech processing, this work can inspire the speech community to explore more on this paradigm.

\begin{figure*}[ht]
    \centering
    \includegraphics[width=0.86\linewidth, height=4.7cm]{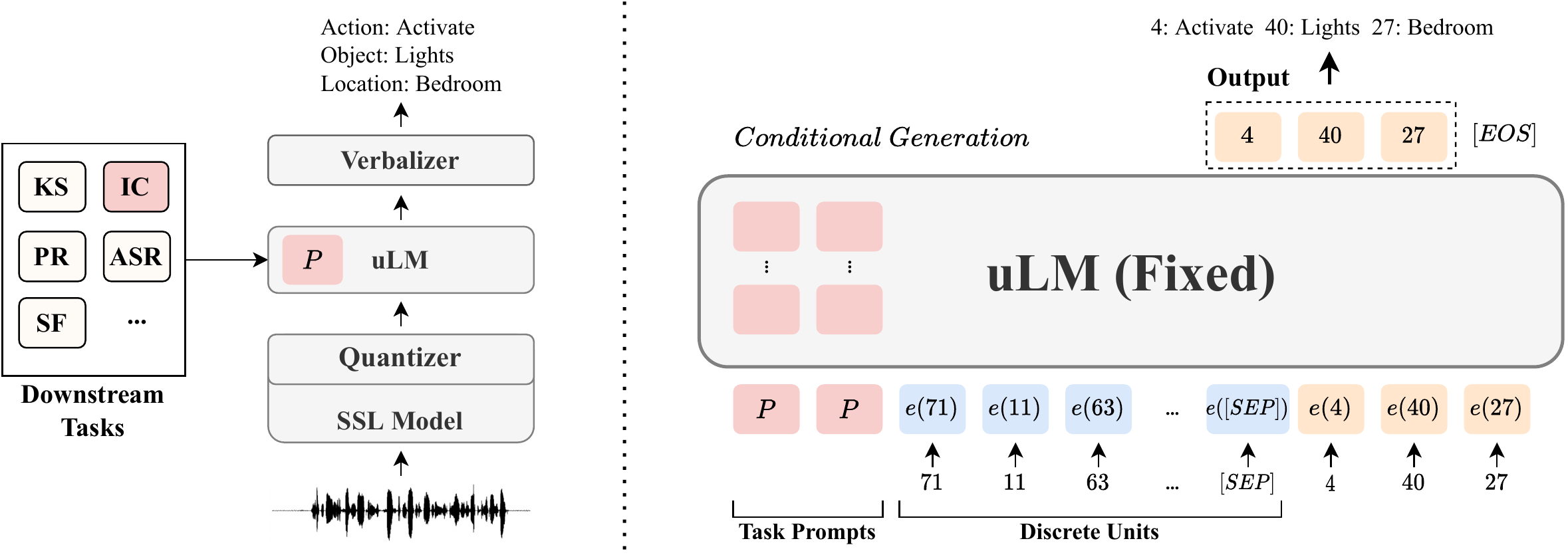}%
    
    \caption{(a) The overview of the proposed framework. Task-specific prompts are applied to the unit language model (uLM) to generate predictions. (b) The uLM performs generation conditioned on the discrete unit sequence and the task prompts.}
    \label{fig:framework}
\end{figure*}

\section{Related Works}
\subsection{SSL Speech Representations}
Learning speech representations with SSL objectives has become a vital research topic in the speech community. 
For example, CPC \cite{DBLP:journals/corr/abs-1807-03748} learns by predicting future features in a contrastive manner.
HuBERT \cite{DBLP:journals/taslp/HsuBTLSM21} maximizes the similarity between output representations and clusters of acoustic features during pre-training. 
To leverage SSL representations, a common way is to build specialized downstream models on top of SSL representations and fine-tune the entire models or only the downstream ones for supervised downstream tasks. 
Based on this, SUPERB \cite{DBLP:conf/interspeech/YangCCLLLLSCLHT21} benchmarks speech SSL techniques with a wide variety of downstream tasks. This work explores an alternative paradigm: we add a fixed, pre-trained LM on top of SSL representations and prompt the LM to generate predictions directly.

\subsection{Generative Spoken Language Model}
Researchers proposed \textbf{Generative Spoken Language Model (GSLM)} \cite{DBLP:journals/corr/abs-2102-01192} to model the rich expressiveness of spoken language based on discovered units of raw audio without any text or labels. 
GSLM first leverages the representation learning ability of SSL speech models and encodes raw speech into a sequence of discrete units by an SSL model and the K-means clustering algorithm. 
A generative \textbf{unit language model (uLM)}, the core component of GSLM, is then trained to perform speech generation on top of discrete units.
The technique shows competitive performance to generate novel speech unconditionally or conditioned on a speech segment.
GSLM can be considered the speech version of GPT-3 \cite{DBLP:journals/corr/abs-2109-07684} and is the first generative speech LM trained on a large speech corpus. The work opens the door to applying prompt tuning methods for speech processing tasks.

\subsection{Prompting and Reprogramming}
Prompting refers to techniques for finding task-specific instructions or templates that steer a pre-trained LM without modifying its parameters \cite{DBLP:journals/corr/abs-2107-13586}.
By concatenating examples and a task description to the original sentence as the input,
GPT-3 \cite{DBLP:journals/corr/abs-2109-07684} performs \emph{in-context learning} to directly generate labels of the sentence.
Although in-context learning yields competitive results, it requires heavy hand-crafted prompt engineering, and it is difficult to scale to smaller pre-trained models \cite{DBLP:conf/naacl/SchickS21}.
To avoid hand-crafted prompt engineering, automatically generating templates \cite{DBLP:conf/emnlp/ShinRLWS20, DBLP:journals/jmlr/RaffelSRLNMZLL20} is another research direction.

Under the premise of fixing pre-trained LMs' parameters, researchers further explore \emph{prompt tuning}, where continuous prompts are learned in the model's embedding space. For example, 
\cite{DBLP:conf/emnlp/LesterAC21, DBLP:journals/corr/abs-2103-10385} learn continuous prompts in LMs' input embedding space. We refer to this kind of methods as \emph{input prompt tuning}.
Another similar technique to input prompt tuning is \emph{model reprogramming} \cite{DBLP:conf/iclr/ElsayedGS19}, where an input transformation function is learned to reprogram a pre-trained model to perform a target task. 
\cite{DBLP:conf/icml/YangTC21, DBLP:journals/corr/abs-2110-03894} have explored reprogramming acoustic models while focusing on single-label classification tasks with a supervised model.
In this work, the proposed framework can perform various speech processing tasks and is not limited to input transformation.
Alternatively, Prefix-Tuning \cite{DBLP:conf/acl/LiL20} and P-Tuning v2 \cite{DBLP:journals/corr/abs-2110-07602} performs \emph{deep prompt tuning}, in which prefix prompts are further prepended at the input of model's hidden layer.
We mainly utilize deep prompt tuning in this work. 
Meanwhile, we also investigate applying prompts only at the input of the LM for comparing different prompting techniques.

\section{Method}
We propose a prompting framework to adapt GSLM to a given downstream task by conditioning the \textbf{uLM} on task-specific prompts. 
Figure~\ref{fig:framework} illustrates the framework. 
An utterance is first encoded into discrete units by an SSL speech model and a K-means quantizer. The uLM then takes the sequence of units as input and prepends it with task-specific prompts.
We then perform conditional generation with the uLM to output units that will be mapped to task labels with a pre-defined verbalizer. In the following, we describe details in the framework, including applying prompts at uLM, controlling the output of conditional generation, and the label mapping with a verbalizer.

\subsection{Prompt Tuning}
\label{sec:prompt_tuning}
A causal uLM $\mathcal{M}$ takes a discrete unit sequence $\bm{u}_x$ as input and autoregressively outputs a sequence $\bm{u}_y = \mathcal{M}(\bm{u)}$ until an end-of-sentence token ``$[EOS]$'' is produced.

In prompt tuning, the parameters of the pre-trained uLM $\mathcal{M}$ are fixed. 
Given $\mathcal{M}$ and a downstream task, a set of trainable task-specific prompt vectors $\mathcal{P}$ is optimized during adaptation with supervision from the task. 
The number of trainable parameters for each task is denoted as $|\mathcal{P}|$.
We adopt deep prompt tuning similar to \cite{DBLP:conf/acl/LiL20, DBLP:journals/corr/abs-2110-07602} in our framework. Given an utterance, the SSL model and the quantizer first encode it into a sequence of discrete units $\bm{u}_x = [u_1, u_2, \cdots, u_T], u_i \in \mathcal{U}$, where $T$ is the unit sequence length, and $\mathcal{U}$ is the unit space of the uLM \footnote{Following GSLM \cite{DBLP:journals/corr/abs-2102-01192}, unit deduplication is also applied universally. (e.g. the unit sequence 71 11 11 63 63 63 becomes 71 11 63.)}.  
Trainable continuous prompts are then applied to (a) the input of the uLM in its embedding space and (b) the input of the attention mechanism in each Transformer block.\\
\noindent\textbf{(a) Prompts at the input of the uLM}\\
Given an unit sequence $\bm{u}_x$ as the original input, the input embedding layer of the uLM $\bm{e}(\cdot): \mathcal{R} \mapsto \mathcal{R}^d$ first transforms it into a sequence of embedding vectors: $\bm{e}(\bm{u}_x) = [\bm{e}(u_1), \bm{e}(u_2), \cdots, \bm{e}(u_T)].$
The sequence is then prepended with continuous prompts and fed into the uLM:
\begin{align}
    [\bm{p}_1^I, \bm{p}_2^{I}, \cdots, \bm{p}_l^{I}, \bm{e}(\bm{u}_x)]
\end{align}
where $\bm{p}^I$ are trainable vectors in $\mathcal{P}$, and $l$ is the prompt length.\\
\noindent\textbf{(b) Prompts at the input of the attention mechanism}\\
Solely applying prompts to the input embedding may not be powerful enough to steer a pre-trained LM \cite{DBLP:journals/corr/abs-2110-07602}. 
Therefore, we also apply prompts to the input of the self-attention mechanism \cite{DBLP:conf/nips/VaswaniSPUJGKP17} in every Transformer block of the uLM.
Given a Transformer block that takes the embedding $\textbf{x}=[\textbf{x}_1,\textbf{x}_2,\cdots,\textbf{x}_T]$ as input, we manipulate the key $K$ and value $V$ in the attention function $Attn(Q, K, V)$ \cite{DBLP:conf/nips/VaswaniSPUJGKP17}: 
\begin{align}
K &= Concat(\bm{p}^K,\bm{x}_{l+1:T})W^K\\
V &= Concat(\bm{p}^V,\bm{x}_{l+1:T})W^V
\end{align}
where $\bm{p}^K$ and $\bm{p}^V$ are trainable vectors in $\mathcal{P}$. 
That is, we replace the first $l$ vectors of $\bm{x}$ with the trainable prompt vectors.

\subsection{Conditional Generation and Verbalizer}
\label{sec:cond_gen}
To leverage the generation capability of the uLM for inference in downstream tasks, we reformulate all tasks into conditional generation problems.
The uLM generates an output sequence $\bm{u}_y$ conditioned on the input unit sequence $\bm{u}_x$ and task-specific prompts $\mathcal{P}$. 
Let $\bm{y} = (y_1,...y_{n})$, $y \in \mathcal{Y}$, be a sequence of $n$ task labels and $ \mathcal{Y}$ is the label space of a task.
The task label length $|\bm{y}|$ is flexible depending on the task. 
For example, in classification tasks, $\bm{y}$ can be a single label or multiple labels. In recognition tasks, $\bm{y}$ can be a character sequence. 

To connect the LM's output with the labels of downstream tasks, a
\emph{verbalizer} \cite{DBLP:conf/naacl/SchickS21, DBLP:conf/eacl/SchickS21} is introduced in the paradigm of prompting LMs.
The verbalizer is a one-to-one mapping $v: \mathcal{Y} \mapsto \mathcal{U}$ that maps from the task label space to vocabulary of the language model. In the case of the uLM, the vocabulary is the units $\mathcal{U}$.
With the help of the verbalizer, the output units can be mapped back to task labels.
For example, in Intent Classification, the output units $\bm{u}_y = [4, 40, 27]$ can be interpreted as intent labels [``Active'', ``Lights'', ``Bedroom''].

The trainable prompts are then optimized with a loss function $\mathcal{L}$, which is cross-entropy for every task in this work:
\begin{align}
    \mathcal{P} = \mathop{\arg\min}_{\mathcal{P}} \mathcal{L}(\mathcal{M}(\mathcal{P}, \bm{u}_x), v(\bm{y}))
\end{align}

\section{Experiment Setup}
\subsection{Tasks and Datasets}
We evaluate the proposed framework on various speech processing tasks, including speech classification tasks: Keyword Spotting (KS) and Intent Classification (IC); sequence generation tasks: ASR and Slot Filling (SF). Table \ref{tab:tasks} gives a brief summary for each task. 
We follow the same dataset and data splits as in SUPERB \cite{DBLP:conf/interspeech/YangCCLLLLSCLHT21}. 
Due to space limitations, please refer to SUPERB for more detail descriptions.

\newcommand{\tabincell}[2]{\begin{tabular}{@{}#1@{}}#2\end{tabular}}
\begin{table}[!t]
\footnotesize
\caption{Summary of downstream tasks used in the work. \textbf{SLU} denotes Spoken Language Understanding. \textbf{CLS}: Classification. \textbf{SG}: Sequence Generation. \textbf{$\overline{|\bm{y}|}$}: average label length in the task.}
\centering
    \begin{tabular}{|c|c|c|c|c|c|c}
    \hline
        \multicolumn{2}{|c|}{Task} & Type  & $N_{class}$  & $\overline{|\bm{y}|}$& Dataset\\
        \hline
        \hline
         KS & Detection & CLS & 12  & 1 &\cite{DBLP:journals/corr/abs-1804-03209}\\
         \hline
         IC & SLU & CLS & 24  & 3 & \cite{DBLP:conf/interspeech/LugoschRITB19}\\
         \hline
         ASR &  Recognition & SG & 29  & 173 & \cite{DBLP:conf/icassp/PanayotovCPK15}\\
         \hline
         SF  & \tabincell{c}{Recogition + SLU} & SG & 69  & 54 & \cite{DBLP:conf/icassp/LaiCL0G21}\\
         \hline
    \end{tabular}
\label{tab:tasks}
\end{table}

\subsection{Implementation Details}
\noindent\textbf{uLM}
We use the checkpoints of the pre-trained uLMs corresponding to HuBERT and CPC representation with 100 clusters on \emph{fairseq}\footnote{\url{https://github.com/pytorch/fairseq/tree/main/examples/textless_nlp/gslm}}  \cite{ott2019fairseq}.
The uLM is a causal LM consisting of 12 Transformer decoder layers with 151M parameters. All the parameters are fixed during prompt tuning.

\noindent\textbf{Verbalizer}
We utilize a simple frequency-based algorithm to implement the verbalizer, in which no model estimation \cite{DBLP:conf/eacl/SchickS21, DBLP:conf/acl/GaoFC20} is involved to simplify the pipeline.
    For $N$ classes in the task (c.f. Table~\ref{tab:tasks}), we map those $N$ classes into $N$ unique units by the following steps: (1) Find and sort top-$N$ frequent units in the input of the training data, denoted as $[u_1, u_2,...u_N]$. (2) Find and sort top-$N$ frequent classes in the ground truth of the training data, denoted as $[c_1, c_2, ...c_N]$. (3) Define the verbalizer $v$ as an one-to-one function: $v(c_i) = u_i$.
We find that applying the frequency-based verbalizer improves the performance by a small margin compared to random assignment. 
We do not show the experiment result of random assignment due to space limitations.

\noindent\textbf{Prompt Length}
We find that the optimal prompt length varies between tasks. For speech classification tasks, we used as fewer prompts as possible while keeping the performance competitive. Regarding sequence generation tasks, we use prompt length $l = 180$, where 4.5M parameters, which equals 3\% parameters of the uLM, are trainable. 

\section{Results}
\subsection{Speech Classification Tasks}
Table~\ref{tab:exp} shows the result of the proposed prompt tuning (PT) framework in multiple tasks.
For comparison, we also list the performance of fine-tuning the uLM (FT-LM), where the same framework is adopted but with the entire uLM trainable. We also list the performance of fine-tuning the specialized downstream models (FT-DM) as in SUPERB \cite{DBLP:conf/interspeech/YangCCLLLLSCLHT21} as a strong baseline.
SUPERB utilizes linear models as downstream models for KS and IC, and 2-layer Bi-LSTMs for ASR and SF.
As shown in Table~\ref{tab:exp a}, in speech classification tasks, prompt tuning achieves competitive performance with fewer trainable parameters.
Notably, in IC, a multi-label classification task, prompt tuning outperforms fine-tuning the entire uLM or downstream models.
The advantage might be that a sequence generation model is suitable for learning the correlation between labels \cite{DBLP:conf/coling/YangSLMWW18}.
\footnote{For an example in IC, object ``lights" can be ``activated" but cannot be ``decreased."}

\begin{table}[h]
\centering
\caption{Performance of prompt tuning in various speech processing tasks. Fine-tuning baselines are listed for comparison. \textbf{PT}: Prompt Tuning. \textbf{FT-LM}: Fine-Tuning the pre-trained uLM. \textbf{FT-DM}: Fine-Tuning the Downstream Model as in SUPERB.\\ \textbf{\#}: Number of trainable parameters.}
\begin{subtable}{\linewidth}
    \caption{Performance on speech classification tasks.}
    \centering
    \begin{tabular}{crrrrrr}
    \hline
     \multirow{2}{*}{\textbf{Scenarios}}  & \multicolumn{2}{c|}{KS} & \multicolumn{2}{c}{IC} \\
     \cline{2-5}
     {} & Acc $\uparrow$ & \tabincell{c}{\textbf{\#}} & Acc $\uparrow$  & \tabincell{c}{\textbf{\#}} \\
     \hline 
     HuBERT-PT & 95.16 & 0.08M & \textbf{98.40} & 0.15M \\
     FT-LM & 94.03 & 151M & 97.63 & 151M \\
     FT-DM & \textbf{96.30} & 0.2M & 98.34 & 0.2M  \\
     \hline
     \hline
     CPC-PT & \textbf{93.54} & 0.05M & \textbf{97.57} & 0.05M  \\
     FT-LM & 93.48 & 151M & 95.62 & 151M \\
     FT-DM & 91.88 & 0.07M & 64.09 & 0.07M  \\
     
    \hline
\end{tabular}
\label{tab:exp a}
\end{subtable}

\begin{subtable}{\linewidth}
    \centering
    \footnotesize
    \caption{Performance on sequence generation tasks.}
    \begin{tabular}{crrrrr}
        \hline
         \multirow{2}{*}{\textbf{Scenarios}}  & \multicolumn{2}{c|}{ASR} & \multicolumn{2}{c}{SF} & \multirow{2}{*}{\textbf{\#}} \\
         \cline{2-5}
         {} & WER $\downarrow$ & CER $\downarrow$ & F1 $\uparrow$ & CER $\downarrow$ & {} \\
         \hline 
         HuBERT-PT & 34.17 & 26.14  & 66.90 & 59.47 & 4.5M\\
         FT-LM  & 26.19 & 16.80  & 80.58 & 40.15 & 151M \\
         FT-DM & \textbf{6.42} & \textbf{1.48}  & \textbf{88.53} & \textbf{25.20} & 43M \\
         \hline
         \hline
         CPC-PT & 59.41 & 37.12  & 65.25 & 60.84 & 4.5M \\
         FT-LM & 35.61 & 17.90  & \textbf{79.34} & \textbf{42.64} & 151M \\
         FT-DM & \textbf{20.18} & \textbf{5.25}  & 71.19 & 49.91 & 42.5M \\
         \hline
    \end{tabular}
    \label{tab:exp b}
\end{subtable}
\label{tab:exp}
\end{table}

\subsection{Sequence Generation and Curse of Long Sequences}
We further push the limit of the prompting paradigm to perform challenging sequence generation tasks: ASR and SF.
As shown in Table~\ref{tab:exp b}, we find that even fine-tuning the uLM (FT-LM) is not comparable to the performance of fine-tuning the specialized downstream models (FT-DM), where CTC loss and Bi-LSTMs are adopted. 
To better understand the gap and possible mitigations of proposed prompt tuning, we study the correlation between the label length $|\bm{y}|$ (i.e., sequence to be generated) and the character error rates (CERs) of HuBERT-PT and HuBERT-FT-LM in ASR. Figure~\ref{fig:long seq} shows that the performance drops significantly when it comes to long sequences.

\begin{figure}[!t]
    \centering
    \captionsetup[sub]{font=large,labelformat=parens}
    \resizebox{\linewidth}{!}{
        \begin{tikzpicture}
        \pgfplotsset{
          compat = 1.3,
          every axis plot/.append style={line width=1pt, mark size=2pt},
        }
        \begin{groupplot}[
        group style={group name=fig2,group size=2 by 1}
        ]
        \nextgroupplot[
            ybar,
            xtick={0,100,200,300,400,500,600},
            xticklabels={0,100,200,300,400,500,600},
            x tick label style={font=\large,yshift=1.5mm},
            xtick style={/pgfplots/major tick length=-1.5mm},
            y tick label style={font=\large},
            yticklabel pos=left,
            ymin=0, ymax=100,
            xmin=-50, xmax=650,
            minor y tick num = 3,
            width=7.5cm,
            height=5.2cm,
            bar width=8,
            ymajorgrids,
            xlabel={Label length $|\bm{y}|$} ,
            xlabel style={font=\large},
            ylabel style={font=\large},
            ytick={10,30,50,70,90,100},
            yticklabels={10,30,50,70,90,},
            ylabel={CER (\%)},
            label style = {font=\large},
            legend style={font=\large,at={(0.22,0.94)}, anchor=north}
        ]
        
        
        \addplot+ [mark=no, fill=RYB1] coordinates { (50, 10.86) (150, 8.78) (250, 9.36) (350, 84.77) (450, 100.01) (550, 81.98)};        
        \addplot+ [mark=no, fill=RYB2] coordinates { (50, 20.7)  (150, 21.4) (250, 28.6) (350, 52.4) (450, 73.5) (550, 80.8)};
        \legend{FT-LM, PT}
        
        \nextgroupplot[
          xtick={0,1,2,3,4,5},
          xticklabels ={1,2,3,6,30,60},  
          x tick label style={font=\large},
          ytick={92,94,96,98,100},
          y tick label style={font=\large},
          yticklabel pos=right,
          ymin=90, ymax=100,
          xmin=0, xmax=5,
          xlabel={Prompt length $l$},
          xlabel style={font=\large},
          ylabel style={font=\large},
          width=7.5cm,
          height=5.2cm,
          ylabel={Acc (\%)},
          ylabel shift={-5pt},
          legend columns=4,
          legend style={font=\large,at={(0.73,0.22)}, anchor=north},
        ]
        \addlegendentry{KS};
        \addplot [color=blue, solid, mark=square* ,mark options={solid}] table [y=KS, x expr=\coordindex] {\exptwo};
        \addlegendentry{IC};
        \addplot [color=red, solid, mark=* ,mark options=solid] table [y=IC, x expr=\coordindex] {\exptwo};
    
        \end{groupplot}
        \node[text width=7.5cm,align=center,anchor=north] at ([yshift=-10mm]fig2 c1r1.south) {\subcaption{\label{fig2:label_length}}};
        \node[text width=7.5cm,align=center,anchor=north] at ([yshift=-10mm]fig2 c2r1.south) {\subcaption{\label{fig2:prompt_length}}};
        
        \end{tikzpicture}
    }
    \caption{(a) CER of different label length $|\bm{y}|$ intervals when using HuBERT. (b) Accuracy of HuBERT-PT with different prompt length on KS and IC.}
    \label{fig:long seq}
\end{figure}
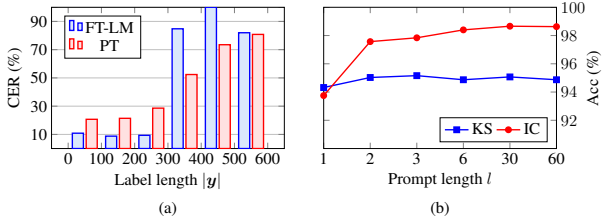

\begin{figure}[!t]
    \centering
        \resizebox{\linewidth}{!}{
            \pgfplotsset{
              compat = 1.3,
              every axis plot/.append style={line width=1pt, mark size=1pt},
              xmode=log,
              xticklabels={$10^3$, $10^4$, $10^5$, $10^6$,$10^7$},
              x tick label style={font=\large},
              ytick={75, 80, 85, 90, 95, 100},
              yticklabels={, 80, 85, 90, 95, 100},
              y tick label style={font=\large},
              ymin=75, ymax=100,
              xmin=1e3, xmax=1e7,
              grid,
              height=5.2cm,
              width=7.5cm,
              label style = {font=\large},
              legend style={font=\large,at={(0.75,0.35)}, anchor=north}
            }
            \begin{tikzpicture}
            \begin{groupplot}[
            group style={group size=2 by 1}
            ]
            \nextgroupplot[
                ylabel={Acc (\%)},
                ylabel shift={-5pt},
                xlabel={Number of trainable parameters $|\mathcal{P}|$},
                xlabel style={at={(1.1,-0.13)},font=\large}
            ]
            \addlegendentry{Input PT};
            \addplot [color=Cerulean, solid, mark=square* ,mark options=solid] table [y=KS, x=param]{\expfour};         
            \addlegendentry{Deep PT};
            \addplot [color=blue, solid, mark=triangle* ,mark options=solid] table [y=KS, x=param] {\expfourDeep};
            \hfill
            \nextgroupplot
            \addlegendentry{Input PT};
            \addplot [color=orange, solid, mark=* ,mark options=solid] table [y=IC, x=param] {\expfour};        
            \addlegendentry{Deep PT};
            \addplot [color=red, solid, mark=* ,mark options=solid] table [y=IC, x=param] {\expfourDeep};
            \end{groupplot}
            \end{tikzpicture}
            
        }
    \caption{Comparison of input prompt tuning (Input PT) and deep prompt tuning (Deep PT). Left: on Keyword Spotting. Right: on Intent Classification.}
    \label{fig:input prompt tuning}
\end{figure}
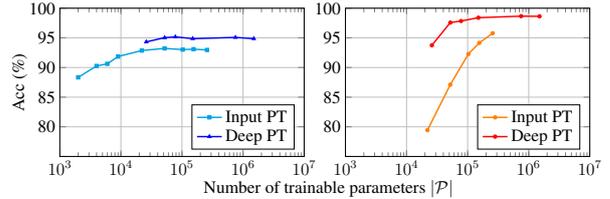

We surmise that the performance drop results from that the uLM is a causal, decoder-only model, which may be unsuitable for recognizing long sequences. Similar phenomena have also been observed in the NLP field.
Due to the limitation of the unidirectional attention mechanism, generative models need more parameters and pre-trained data to work on Natural Language Understanding (NLU) tasks \cite{DBLP:journals/corr/abs-2103-10360}. In a more complex task, text summarization, GPT-2 is also suffered from long sequences \cite{DBLP:conf/acl/LiL20}. Although GPT-3 \cite{DBLP:journals/corr/abs-2109-07684} shows competitive performance by performing NLU tasks as a generation problem, a much larger model (175B parameters) is also required. The uLM has only 151M parameters and therefore falls behind the fine-tuning of specialized downstream models in challenging sequence generation tasks, where label lengths are way longer than those in \cite{DBLP:conf/acl/LiL20} (c.f. Table~\ref{tab:tasks}).
We will continue the discussion of the limitations of existing pre-trained generative models and possible research directions in Section~\ref{discussion}.

\subsection{Prompt Length} \label{exp:prompt length}
We vary the prompt length $l$ in HuBERT-PT to study the effect of the number of trainable parameters.
In Figrue~\ref{fig2:prompt_length}, the result shows that as the prompt length increases, there is a trend of increasing performance.
It is worth noting that it still achieves a reasonable accuracy with prompt length only equal to 2, where 52K trainable parameters are introduced.

\subsection{Input Prompt tuning}
When the inner parameters of pre-trained LMs are not accessible, only input prompt tuning can be applied. 
Thus, we further study the IC and KS performance of our approach when prompts are only used at the input of the uLM. Figure~\ref{fig:input prompt tuning} shows that although deep prompt tuning (i.e. applying prompts further at LMs' hidden layer described in section~\ref{sec:prompt_tuning}(b)) consistently outperforms input prompt tuning, the latter can also achieve competitive performance with sufficient trainable parameters.

\section{Discussion and Future Works} \label{discussion}
Unlike prompt tuning in NLP, the meaning of the uLM's vocabulary is not obvious.
In NLP, it is usually simple to identify how to define the verbalizer \cite{DBLP:conf/eacl/SchickS21}, and often the verbalizer is even an identity function when the prediction target is the vocabulary itself \cite{DBLP:conf/naacl/SchickS21}.
This paper leverages a heuristic, frequency-based approach to define the verbalizer. How to better identify the mapping between discovered units and task labels is critical for performance and remains unsolved.

Although the experiments show that the proposed framework achieves competitive results in speech classification tasks, we are restricted by the nature of the uLM when performing challenging sequence generation tasks.
In NLP, prompting on text classification tasks has also achieved remarkable results \cite{DBLP:journals/corr/abs-2107-13586, DBLP:conf/naacl/SchickS21, DBLP:conf/eacl/SchickS21,  han2021ptr}. 
However, to solve more difficult text generation tasks (e.g. summarization, translation), larger and more powerful pre-trained LMs including Prefix LMs (e.g. UniLMs \cite{DBLP:conf/nips/00040WWLWGZH19, bao2020unilmv2}) and Encoder-Decoder LMs (e.g. T5 \cite{DBLP:journals/jmlr/RaffelSRLNMZLL20}, BART \cite{DBLP:conf/acl/LewisLGGMLSZ20}) are often introduced \cite{DBLP:journals/corr/abs-2107-13586, DBLP:conf/acl/LiL20,  DBLP:journals/corr/abs-2110-07602, qin2022lfpt}.
For speech processing tasks, the problems might be even more difficult since the model is expected to perform recognition (KS, ASR), understanding (IC), or both at the same time (SF), while there are few LMs available for speech.
We hope this work motivates the speech community to invest in diverse and effective speech LMs.

\section{Conclusions}
In this work, we propose a prompt tuning framework based on Generative Spoken Language Model (GSLM) for speech processing tasks.
The experiment results shows that the language model can be guided to directly generate the answers by tuning a limited number of task-specific vectors.
In classification tasks, the framework achieves competitive performance compared to fine-tuning the entire downstream models.
We also investigate the limitation of the framework on challenging sequence generation tasks.
This paper is the first exploration of prompt tuning paradigm for speech processing tasks.
We believe this study can motivate the speech research communities to explore the prompting paradigm more.

\bibliographystyle{IEEEtran}
\bibliography{mybib}

\begin{thebibliography}{10}
\providecommand{\url}[1]{#1}
\csname url@samestyle\endcsname
\providecommand{\newblock}{\relax}
\providecommand{\bibinfo}[2]{#2}
\providecommand{\BIBentrySTDinterwordspacing}{\spaceskip=0pt\relax}
\providecommand{\BIBentryALTinterwordstretchfactor}{4}
\providecommand{\BIBentryALTinterwordspacing}{\spaceskip=\fontdimen2\font plus
\BIBentryALTinterwordstretchfactor\fontdimen3\font minus
  \fontdimen4\font\relax}
\providecommand{\BIBforeignlanguage}[2]{{%
\expandafter\ifx\csname l@#1\endcsname\relax
\typeout{** WARNING: IEEEtran.bst: No hyphenation pattern has been}%
\typeout{** loaded for the language `#1'. Using the pattern for}%
\typeout{** the default language instead.}%
\else
\language=\csname l@#1\endcsname
\fi
#2}}
\providecommand{\BIBdecl}{\relax}
\BIBdecl

\bibitem{DBLP:journals/corr/abs-1807-03748}
A.~van~den Oord, Y.~Li, and O.~Vinyals, ``Representation learning with
  contrastive predictive coding,'' \emph{CoRR}, vol. abs/1807.03748, 2018.

\bibitem{DBLP:conf/nips/BaevskiZMA20}
A.~Baevski, Y.~Zhou, A.~Mohamed, and M.~Auli, ``wav2vec 2.0: {A} framework for
  self-supervised learning of speech representations,'' in \emph{NeurIPS},
  2020.

\bibitem{DBLP:journals/taslp/HsuBTLSM21}
W.~Hsu, B.~Bolte, Y.~H. Tsai, K.~Lakhotia, R.~Salakhutdinov, and A.~Mohamed,
  ``Hubert: Self-supervised speech representation learning by masked prediction
  of hidden units,'' \emph{{IEEE} {ACM} Trans. Audio Speech Lang. Process.},
  vol.~29, pp. 3451--3460, 2021.

\bibitem{chen2021wavlm}
S.~Chen, C.~Wang, Z.~Chen, Y.~Wu, S.~Liu, Z.~Chen, J.~Li, N.~Kanda,
  T.~Yoshioka, X.~Xiao \emph{et~al.}, ``Wavlm: Large-scale self-supervised
  pre-training for full stack speech processing,'' \emph{arXiv preprint
  arXiv:2110.13900}, 2021.

\bibitem{DBLP:conf/icassp/BaevskiM20}
A.~Baevski and A.~Mohamed, ``Effectiveness of self-supervised pre-training for
  {ASR},'' in \emph{{ICASSP}}.\hskip 1em plus 0.5em minus 0.4em\relax {IEEE},
  2020, pp. 7694--7698.

\bibitem{DBLP:journals/corr/abs-2111-09296}
A.~Babu, C.~Wang, A.~Tjandra, K.~Lakhotia, Q.~Xu, N.~Goyal, K.~Singh, P.~von
  Platen, Y.~Saraf, J.~Pino, A.~Baevski, A.~Conneau, and M.~Auli, ``{XLS-R:}
  self-supervised cross-lingual speech representation learning at scale,''
  \emph{CoRR}, vol. abs/2111.09296, 2021.

\bibitem{DBLP:conf/icassp/LaiCL0G21}
C.~Lai, Y.~Chuang, H.~Lee, S.~Li, and J.~R. Glass, ``Semi-supervised spoken
  language understanding via self-supervised speech and language model
  pretraining,'' in \emph{{ICASSP}}.\hskip 1em plus 0.5em minus 0.4em\relax
  {IEEE}, 2021, pp. 7468--7472.

\bibitem{DBLP:conf/interspeech/LinLCL21}
J.~Lin, Y.~Y. Lin, C.~Chien, and H.~Lee, ``{S2VC:} {A} framework for any-to-any
  voice conversion with self-supervised pretrained representations,'' in
  \emph{Interspeech}.\hskip 1em plus 0.5em minus 0.4em\relax {ISCA}, 2021, pp.
  836--840.

\bibitem{baevski2021unsupervised}
A.~Baevski, W.-N. Hsu, A.~Conneau, and M.~Auli, ``Unsupervised speech
  recognition,'' \emph{Advances in Neural Information Processing Systems},
  vol.~34, 2021.

\bibitem{DBLP:conf/interspeech/YangCCLLLLSCLHT21}
S.~Yang, P.~Chi, Y.~Chuang, C.~J. Lai, K.~Lakhotia, Y.~Y. Lin, A.~T. Liu,
  J.~Shi, X.~Chang, G.~Lin, T.~Huang, W.~Tseng, K.~Lee, D.~Liu, Z.~Huang,
  S.~Dong, S.~Li, S.~Watanabe, A.~Mohamed, and H.~Lee, ``{SUPERB:} speech
  processing universal performance benchmark,'' in \emph{Interspeech}.\hskip
  1em plus 0.5em minus 0.4em\relax {ISCA}, 2021, pp. 1194--1198.

\bibitem{Tsai2022SUPERBSGES}
H.-S. Tsai, H.-J. Chang, W.-C. Huang, Z.~Huang, K.~Lakhotia, S.~wen Yang,
  S.~Dong, A.~T. Liu, C.-I.~J. Lai, J.~Shi, X.~Chang, P.~Hall, H.-J. Chen,
  S.-W. Li, S.~Watanabe, A.~rahman Mohamed, and H.~yi~Lee, ``Superb-sg:
  Enhanced speech processing universal performance benchmark for semantic and
  generative capabilities,'' 2022.

\bibitem{lai2021parp}
C.-I.~J. Lai, Y.~Zhang, A.~H. Liu, S.~Chang, Y.-L. Liao, Y.-S. Chuang, K.~Qian,
  S.~Khurana, D.~Cox, and J.~Glass, ``Parp: Prune, adjust and re-prune for
  self-supervised speech recognition,'' \emph{Advances in Neural Information
  Processing Systems}, vol.~34, 2021.

\bibitem{DBLP:journals/corr/abs-2107-13586}
P.~Liu, W.~Yuan, J.~Fu, Z.~Jiang, H.~Hayashi, and G.~Neubig, ``Pre-train,
  prompt, and predict: {A} systematic survey of prompting methods in natural
  language processing,'' \emph{CoRR}, vol. abs/2107.13586, 2021.

\bibitem{DBLP:conf/acl/LiL20}
X.~L. Li and P.~Liang, ``Prefix-tuning: Optimizing continuous prompts for
  generation,'' in \emph{{ACL/IJCNLP} {(1)}}.\hskip 1em plus 0.5em minus
  0.4em\relax Association for Computational Linguistics, 2021, pp. 4582--4597.

\bibitem{DBLP:journals/corr/abs-2103-10385}
X.~Liu, Y.~Zheng, Z.~Du, M.~Ding, Y.~Qian, Z.~Yang, and J.~Tang, ``{GPT}
  understands, too,'' \emph{CoRR}, vol. abs/2103.10385, 2021.

\bibitem{DBLP:journals/corr/abs-2110-07602}
X.~Liu, K.~Ji, Y.~Fu, Z.~Du, Z.~Yang, and J.~Tang, ``P-tuning v2: Prompt tuning
  can be comparable to fine-tuning universally across scales and tasks,''
  \emph{CoRR}, vol. abs/2110.07602, 2021.

\bibitem{DBLP:conf/emnlp/LesterAC21}
B.~Lester, R.~Al{-}Rfou, and N.~Constant, ``The power of scale for
  parameter-efficient prompt tuning,'' in \emph{{EMNLP} {(1)}}.\hskip 1em plus
  0.5em minus 0.4em\relax Association for Computational Linguistics, 2021, pp.
  3045--3059.

\bibitem{DBLP:journals/corr/abs-2102-01192}
K.~Lakhotia, E.~Kharitonov, W.~Hsu, Y.~Adi, A.~Polyak, B.~Bolte, T.~A. Nguyen,
  J.~Copet, A.~Baevski, A.~Mohamed, and E.~Dupoux, ``Generative spoken language
  modeling from raw audio,'' \emph{CoRR}, vol. abs/2102.01192, 2021.

\bibitem{DBLP:journals/corr/abs-2109-07684}
G.~I. Winata, A.~Madotto, Z.~Lin, R.~Liu, J.~Yosinski, and P.~Fung, ``Language
  models are few-shot multilingual learners,'' \emph{CoRR}, vol.
  abs/2109.07684, 2021.

\bibitem{DBLP:conf/naacl/SchickS21}
T.~Schick and H.~Sch{\"{u}}tze, ``It's not just size that matters: Small
  language models are also few-shot learners,'' in \emph{{NAACL-HLT}}.\hskip
  1em plus 0.5em minus 0.4em\relax Association for Computational Linguistics,
  2021, pp. 2339--2352.

\bibitem{DBLP:conf/emnlp/ShinRLWS20}
T.~Shin, Y.~Razeghi, R.~L.~L. IV, E.~Wallace, and S.~Singh, ``Autoprompt:
  Eliciting knowledge from language models with automatically generated
  prompts,'' in \emph{{EMNLP} {(1)}}.\hskip 1em plus 0.5em minus 0.4em\relax
  Association for Computational Linguistics, 2020, pp. 4222--4235.

\bibitem{DBLP:journals/jmlr/RaffelSRLNMZLL20}
C.~Raffel, N.~Shazeer, A.~Roberts, K.~Lee, S.~Narang, M.~Matena, Y.~Zhou,
  W.~Li, and P.~J. Liu, ``Exploring the limits of transfer learning with a
  unified text-to-text transformer,'' \emph{J. Mach. Learn. Res.}, vol.~21, pp.
  140:1--140:67, 2020.

\bibitem{DBLP:conf/iclr/ElsayedGS19}
G.~F. Elsayed, I.~J. Goodfellow, and J.~Sohl{-}Dickstein, ``Adversarial
  reprogramming of neural networks,'' in \emph{{ICLR} (Poster)}.\hskip 1em plus
  0.5em minus 0.4em\relax OpenReview.net, 2019.

\bibitem{DBLP:conf/icml/YangTC21}
C.~H. Yang, Y.~Tsai, and P.~Chen, ``Voice2series: Reprogramming acoustic models
  for time series classification,'' in \emph{{ICML}}, ser. Proceedings of
  Machine Learning Research, vol. 139.\hskip 1em plus 0.5em minus 0.4em\relax
  {PMLR}, 2021, pp. 11\,808--11\,819.

\bibitem{DBLP:journals/corr/abs-2110-03894}
H.~Yen, P.~Ku, C.~H. Yang, H.~Hu, S.~M. Siniscalchi, P.~Chen, and Y.~Tsao, ``A
  study of low-resource speech commands recognition based on adversarial
  reprogramming,'' \emph{CoRR}, vol. abs/2110.03894, 2021.

\bibitem{DBLP:conf/nips/VaswaniSPUJGKP17}
A.~Vaswani, N.~Shazeer, N.~Parmar, J.~Uszkoreit, L.~Jones, A.~N. Gomez,
  L.~Kaiser, and I.~Polosukhin, ``Attention is all you need,'' in
  \emph{{NIPS}}, 2017, pp. 5998--6008.

\bibitem{DBLP:conf/eacl/SchickS21}
T.~Schick and H.~Sch{\"{u}}tze, ``Exploiting cloze-questions for few-shot text
  classification and natural language inference,'' in \emph{{EACL}}.\hskip 1em
  plus 0.5em minus 0.4em\relax Association for Computational Linguistics, 2021,
  pp. 255--269.

\bibitem{DBLP:journals/corr/abs-1804-03209}
P.~Warden, ``Speech commands: {A} dataset for limited-vocabulary speech
  recognition,'' \emph{CoRR}, vol. abs/1804.03209, 2018.

\bibitem{DBLP:conf/interspeech/LugoschRITB19}
L.~Lugosch, M.~Ravanelli, P.~Ignoto, V.~S. Tomar, and Y.~Bengio, ``Speech model
  pre-training for end-to-end spoken language understanding,'' in
  \emph{{INTERSPEECH}}.\hskip 1em plus 0.5em minus 0.4em\relax {ISCA}, 2019,
  pp. 814--818.

\bibitem{DBLP:conf/icassp/PanayotovCPK15}
V.~Panayotov, G.~Chen, D.~Povey, and S.~Khudanpur, ``Librispeech: An {ASR}
  corpus based on public domain audio books,'' in \emph{{ICASSP}}.\hskip 1em
  plus 0.5em minus 0.4em\relax {IEEE}, 2015, pp. 5206--5210.

\bibitem{ott2019fairseq}
M.~Ott, S.~Edunov, A.~Baevski, A.~Fan, S.~Gross, N.~Ng, D.~Grangier, and
  M.~Auli, ``fairseq: A fast, extensible toolkit for sequence modeling,'' in
  \emph{Proceedings of NAACL-HLT 2019: Demonstrations}, 2019.

\bibitem{DBLP:conf/acl/GaoFC20}
T.~Gao, A.~Fisch, and D.~Chen, ``Making pre-trained language models better
  few-shot learners,'' in \emph{{ACL/IJCNLP} {(1)}}.\hskip 1em plus 0.5em minus
  0.4em\relax Association for Computational Linguistics, 2021, pp. 3816--3830.

\bibitem{DBLP:conf/coling/YangSLMWW18}
P.~Yang, X.~Sun, W.~Li, S.~Ma, W.~Wu, and H.~Wang, ``{SGM:} sequence generation
  model for multi-label classification,'' in \emph{{COLING}}.\hskip 1em plus
  0.5em minus 0.4em\relax Association for Computational Linguistics, 2018, pp.
  3915--3926.

\bibitem{DBLP:journals/corr/abs-2103-10360}
Z.~Du, Y.~Qian, X.~Liu, M.~Ding, J.~Qiu, Z.~Yang, and J.~Tang, ``All {NLP}
  tasks are generation tasks: {A} general pretraining framework,'' \emph{CoRR},
  vol. abs/2103.10360, 2021.

\bibitem{han2021ptr}
X.~Han, W.~Zhao, N.~Ding, Z.~Liu, and M.~Sun, ``Ptr: Prompt tuning with rules
  for text classification,'' \emph{arXiv preprint arXiv:2105.11259}, 2021.

\bibitem{DBLP:conf/nips/00040WWLWGZH19}
L.~Dong, N.~Yang, W.~Wang, F.~Wei, X.~Liu, Y.~Wang, J.~Gao, M.~Zhou, and
  H.~Hon, ``Unified language model pre-training for natural language
  understanding and generation,'' in \emph{NeurIPS}, 2019, pp.
  13\,042--13\,054.

\bibitem{bao2020unilmv2}
H.~Bao, L.~Dong, F.~Wei, W.~Wang, N.~Yang, X.~Liu, Y.~Wang, J.~Gao, S.~Piao,
  M.~Zhou \emph{et~al.}, ``Unilmv2: Pseudo-masked language models for unified
  language model pre-training,'' in \emph{International Conference on Machine
  Learning}.\hskip 1em plus 0.5em minus 0.4em\relax PMLR, 2020, pp. 642--652.

\bibitem{DBLP:conf/acl/LewisLGGMLSZ20}
M.~Lewis, Y.~Liu, N.~Goyal, M.~Ghazvininejad, A.~Mohamed, O.~Levy, V.~Stoyanov,
  and L.~Zettlemoyer, ``{BART:} denoising sequence-to-sequence pre-training for
  natural language generation, translation, and comprehension,'' in
  \emph{{ACL}}.\hskip 1em plus 0.5em minus 0.4em\relax Association for
  Computational Linguistics, 2020, pp. 7871--7880.

\bibitem{qin2022lfpt}
C.~Qin and S.~Joty, ``{LFPT}5: A unified framework for lifelong few-shot
  language learning based on prompt tuning of t5,'' in \emph{International
  Conference on Learning Representations}, 2022.

\end{thebibliography}
\end{document}